\documentclass[a4paper,11pt]{article}
\usepackage{lineno}
\usepackage{enumitem}
\usepackage{float}
\usepackage{graphicx}
\usepackage{subcaption}
\usepackage[font=small,labelfont=bf,
   justification=justified,
   format=plain]{caption} % 'format=plain' avoids hanging indentation
\pdfoutput=1 % if your are submitting a pdflatex (i.e. if you have
\usepackage{jinstpub} % for details on the use of the package, please
\usepackage[sorting=none]{biblatex}

\usepackage{tabu}
\usepackage{notoccite}
\DeclareMathOperator\erf{erf}
\DeclareMathOperator\Binom{Binom}
\DeclareMathOperator\Normal{Normal}
\DeclareMathOperator\SkewNormal{SkewNormal}

\title{FlameNEST: Explicit Profile Likelihoods with the Noble Element Simulation Technique}

\author[a,1]{R. S. James,\note{Corresponding author.}}
\author[b,1]{J. Palmer,}
\author[b,c]{A. Kaboth,}
\author[a]{C. Ghag,}
\author[d,e]{and J. Aalbers}

\affiliation[a]{University College London,\\Gower St, London, United Kingdom}
\affiliation[b]{Royal Holloway University of London,\\Egham Hill, Egham, United Kingdom}
\affiliation[c]{Rutherford Appleton Laboratory, \\Didcot, United Kingdom}
\affiliation[d]{Stanford University, \\Stanford, CA 94305-4085 USA}
\affiliation[e]{SLAC National Accelerator Laboratory, \\Menlo Park, CA 94025-7015, USA}
\emailAdd{robert.james.19@ucl.ac.uk}
\emailAdd{jordan.palmer.2014@live.rhul.ac.uk}

\abstract{We present FlameNEST, a framework providing explicit likelihood evaluations in noble element particle detectors using data-driven models from the Noble Element Simulation Technique. FlameNEST provides a way to perform statistical analyses on real data with no dependence on large, computationally expensive Monte Carlo simulations by evaluating the likelihood on an event-by-event basis using analytic probability elements convolved together in a single TensorFlow multiplication. Furthermore, this robust framework creates opportunities for simple inter-collaborative analyses which will be fundamental for the future of experimental dark matter physics.}


\begin{document}
\maketitle
\flushbottom

\section{Introduction}
\label{sec:intro}

Observations on both galactic and cosmological scales have found that dark matter constitutes approximately $85\%$ of the matter density in the universe \cite{Garrett_2011, 2020planck}. Over the past decade, time projection chambers (TPCs) containing liquefied noble elements have become the leading technology in the search for the medium of dark matter \cite{LUX_results, X1T_results, LZ_sense, XnT_sens}. Rare event searches such as these often choose to use frequentist hypothesis testing to present their results \cite{2021statswhite}. The central object of such tests is the likelihood which may be obtained via computation of a differential event rate $R^j(\{O_i\})$. This is the number of expected events from the $j^{th}$ signal or background source producing a given set of observables $\{O_i\}$, when integrated over observable space. Experiments today estimate such differential event rates by filling multi-dimensional histograms (templates) in a binned space of observables using Monte Carlo (MC) techniques. Underlying `nuisance' parameters may be incorporated by creating multiple templates and interpolating between them -- these are parameters which influence the event probability model but are of secondary interest to the experiment. Filling these templates to the requisite statistical accuracy scales exponentially with both the number of observables and the number of nuisance parameters, making such analyses computationally unwieldy. A common compromise is to restrict the number of observables and limit the number of underlying nuisance parameters, the former reducing the signal/background discrimination of the detector and the latter making the analysis less robust.

Flamedisx is an open-source Python package allowing for likelihood evaluation scaling approximately linearly rather than exponentially in the number of nuisance parameters. Further to this, there is no scaling with the inclusion of certain additional observables, making the inclusion of many more such dimensions computationally feasible \cite{Aalbers_2020}. This is achieved by calculating likelihoods on an event-by-event basis using real experimental data. Flamedisx computes a sum over `hidden variables' where each term is a product of conditional probabilities calculated from the analytic probability density/mass function (PDF/PMF) of one part of the detector response model -- the distinction here comes in modelling continuous versus discrete variables. The computation is performed using \texttt{TensorFlow} \cite{tensorflow_developers_2021_5095721}, which allows for automatic differentiation to facilitate likelihood maximisation. \texttt{TensorFlow} is greatly accelerated when run on a GPU, increasing computation speed roughly hundred-fold in the case of Flamedisx.

The detector response models originally implemented within Flamedisx, as described in \cite{Aalbers_2020}, are inspired primarily by the XENON1T detector \cite{Aprile_2017}. To extend the Flamedisx framework to be more detector-agnostic, we have incorporated the xenon models of the Noble Element Simulation Technique (NEST) into Flamedisx. NEST is a precise, detector-agnostic parameterisation of excitation, ionisation, and the corresponding scintillation and electroluminescence processes in liquid noble elements as a function of both energy and electric field \cite{szydagis_m_2021_5080263}. These models are constantly being scrutinised and validated against real data collected by a variety of world-leading noble element experiments. In addition to improving the accuracy of and extending the reach of analyses done using Flamedisx, we believe that using the community's gold-standard collection of noble element response models encapsulated in NEST will allow for Flamedisx to be used for future inter-collaborative data analyses between different noble element experiments, further extending physics reach.

This paper outlines the technical challenges of incorporating the NEST models into Flamedisx, a framework henceforth referred to as FlameNEST. We also present the results of a series of validations and discuss the resulting speed implications of our work. The focus throughout will be on dual-phase liquid xenon (LXe) TPCs; however, NEST contains additional models for single-phase gaseous xenon detectors along with liquid argon detectors, incorporation of which into FlameNEST is a future goal.

\section{Dual-phase Liquid Xenon Time Projection Chambers}
\label{sec:TPCs}

\begin{figure}[H]
    \centering
    \includegraphics[width=8cm]{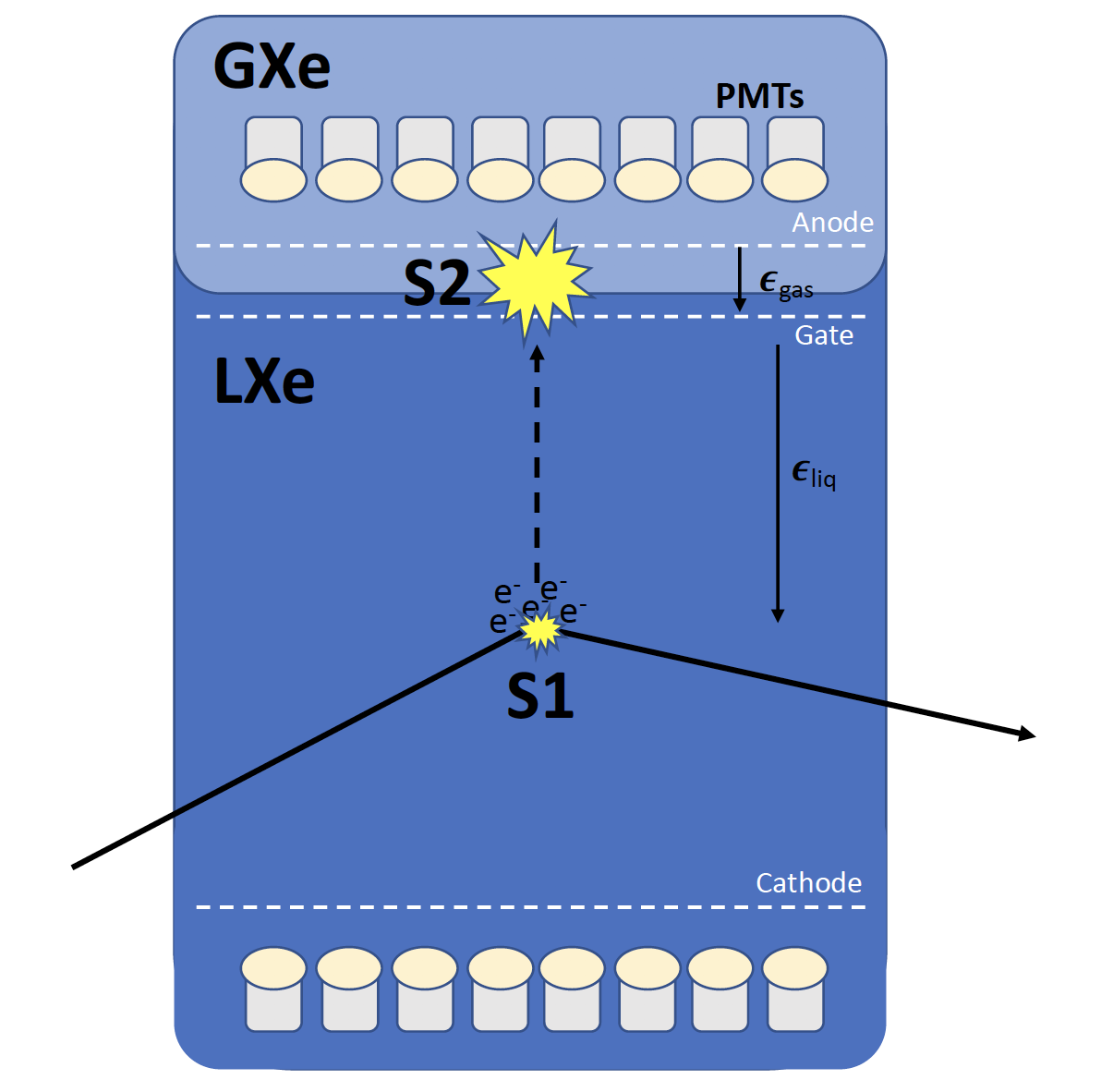}
    \caption{Schematic of a dual-phase LXe TPC showing the signal processes from an interaction in the detector.}
    \label{fig:TPCschematic}
\end{figure}

A general schematic of a dual-phase LXe TPC is shown in Figure \ref{fig:TPCschematic}. These detectors are typically cylindrical and filled with LXe with a thin layer of gaseous xenon (GXe) above. When a particle interacts with an atomic electron or nucleus, xenon dimers are formed, the decay of which produce UV light. This scintillation light is observed by photosensors, typically photomultiplier tubes (PMTs), giving what is denoted the S1 signal of an event. The PMTs are typically arranged in two arrays, located at the top and bottom of the detector. When photons hit a PMT, there is an average probability of producing two electrons at the photocathode (photoelectrons). This is known as the double photoelectron (DPE) effect and must be accounted for when modelling the detector response~\cite{Faham_2015}.

Interactions can additionally produce electron-ion pairs. To detect these ionisation electrons, an electric field $\epsilon_{liq}$ is applied across the LXe bulk. The electrons drift in this field towards the liquid-gas interface where they experience a much larger electric field, $\epsilon_{\text{gas}}$. The higher field is responsible for extracting the electrons from the liquid phase to the gas phase where the electrons can undergo electroluminescence. The secondary photons produced in this process are detected by the same set of PMTs to give what we denote the S2 signal of an event. Some electrons may be absorbed onto impurities within the LXe before reaching the liquid-gas interface. This can be quantified by the electron lifetime, which reduces the average size of the S2 signal towards the bottom of the detector.

The distribution of S2 photons provides $(x,y)$ position information in the radial plane of the detector. The time difference between the S1 and S2 signals, coupled with the electron drift velocity, then allows for reconstruction of the vertical $z$ coordinate. This gives the full set of observables of an interaction event as (S1,S2,$x$,$y$,$z$,$t$), where $t$ is the time at the start of the event.

The relative size of the S1 and S2 signals provides information on the underlying interaction type of the event. Signal and background sources of interest in rare event searches can be classed as inducing either nuclear recoil (NR) or electronic recoil (ER) interactions. For the same energy, NR interactions produce smaller S2s and larger S1s; therefore, the ratio of the two is used as a discrimination metric. To overcome the aforementioned difficulties in filling high-dimensional Monte Carlo templates, current statistical analyses typically opt to divide out position dependence of the S1/S2 values, normalising them to a reference position in the detector. Detector conditions such as temperature and electric field, which can vary throughout the lifetime of an experiment, are typically taken to be constant and data during periods of fluctuation discarded. Likelihood evaluations using Monte Carlo templates often neglect position and time dependence in certain signal and background sources. This reduces the dimensionality of the observable space from (S1,S2,$x$,$y$,$z$,$t$) to `corrected' S1 and S2 values, (S1$c$,S2$c$).

A significant drawback of such a dimensionality reduction is that signal/background discrimination is reduced. This is particularly the case towards the top of the detector, where S2 signals are large and the relative fluctuations in the inferred charge yield are smaller. Thus, a dimensionality reduction reduces the ER/NR discrimination in certain regions of the detector. Furthermore, not correctly accounting for the spatial and temporal dependence of the interaction rates of relevant sources further reduces signal/background discrimination.

The probability distributions describing each stage of this detector response have parameters which are often functions of many other underlying nuisance parameters -- these are specific to the models of the different physical processes constituting the detector response. Whilst auxiliary measurements can constrain them to some degree, a truly robust analysis will allow them to float during inference. Enabling this with a Monte Carlo template likelihood evaluation would lead to exponential scaling in the template generation as more nuisance parameters are included, whereas the Flamedisx computation scales instead approximately linearly with nuisance parameters.

\section{Technical Implementation}

NEST fully models the production of ionised and excited xenon atoms (ions/excitons), including recombination fluctuations, which is subsequently used in modelling the ionisation electron and scintillation photon yields. In contrast, the original Flamedisx models did not feature this extra degree of freedom -- electron/photon production was modelled by smearing the interaction rate spectrum to form an intermediate variable, with the quanta production parameterised in terms of this variable. Additionally, the detector response models translating produced quanta distributions into observable signal distributions in NEST feature a number of extra steps compared to the original Flamedisx models. Consequently, it was not possible to incorporate the NEST models directly into the original tensor structure of Flamedisx. Therefore, the underlying tensor structure of Flamedisx was extended to incorporate the NEST models in full generality. In this section we outline this new structure.

\subsection{Block structure}
\label{subsec:blockStructure}

\begin{figure}[h!]
    \centering
    \includegraphics[width=16cm]{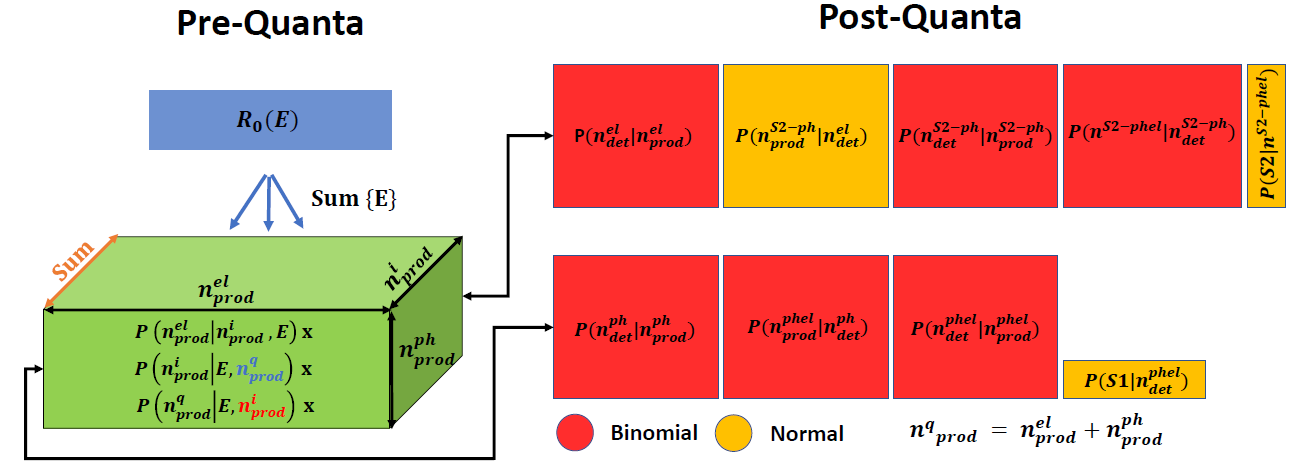}
    \caption{FlameNEST block structure. The blocks are categorised by whether they model pre-quanta processes (production of electrons and photons from an energy deposition) or post-quanta processes (detection of quanta and translation to final signals). The dimensions of each block are indicated graphically. Every block has an additional dimension, not depicted here, over events within a computation batch. The probability distributions for the post-quanta blocks are indicated by their colour -- see Section \ref{subsubsec:preQuanta} for details of the pre-quanta distributions. In the green pre-quanta block, the colour of the text indicates variables that are used for ER (blue) or NR (red) only.}
    \label{fig:newblockstructure}
\end{figure}

The FlameNEST block structure is shown in Figure \ref{fig:newblockstructure}, which may be compared with the original Flamedisx block structure in Figure 3 of \cite{Aalbers_2020}. The pre-quanta stage maps between the interaction rate spectrum of the $j^{th}$ source, $R^j(E,x,y,z,t)$, where $E$ is the energy of the interaction, and produced quanta (photon/electron) distributions. The post-quanta stage maps between these produced quanta distributions and the distributions of signals, S1 and S2. The models depend on the type of interaction of the source with the xenon atoms -- whether an ER or NR occurs. Variables in blue in the figure are only used in place for ER sources, while red variables are only used for NR sources. Event position and time additionally enter at the level of the model functions used in the probabilistic detector response model.

% of the blocks are represented by their line colour: orange depicts blocks with 3 dimensions plus a 4th representing the number of events, whilst black depicts blocks with 2 dimensions with a 3rd representing the number of events.
As outlined in \cite{Aalbers_2020}, Flamedisx computes bounds on any non-observable dimensions of the blocks for each observed event. Each block then has (conditional) probability elements evaluated within those bounds, based on some probability distribution and model functions determining its parameters. The blocks are then multiplied together for different values of energy $E$, multiplied by $R^j(E)$ and the results summed together. In FlameNEST, this sum has the following form:

\begin{equation}
\label{eq:fullSum}
    \sum_{E,e,\gamma,i,j,k,l,m,n,...} P(S1|i) P(i|j) P(j|...) ... P(k|\gamma) P(e,\gamma|E) R^j(E) P(l|e) ... P(m|...) P(n|m) P(S2|n).
\end{equation}

\noindent By evaluating this sum, we obtain the differential event rate $R^j(S1,S2,x,y,z,t)$. Here, $e$ and $\gamma$ are hidden variables representing the number of produced electrons and photons respectively, whilst $i, j, k, l, m, n, ...$ represent other hidden variables in the detector response model such as the number of electrons/photons detected, for example. The bounds are chosen such that each computed probability element will contribute non-negligibly to the sum.

It should be noted that, in some places, NEST uses continuous distributions to model discrete variables, rounding each sampled value during MC simulation. This choice means that the FlameNEST computation needs to include a continuity correction: instead of evaluating $P(X=x)$, we evaluate

\begin{equation}
\label{eq:contCorr}
    P(X \leq x+0.5) - P(X \leq x-0.5).
\end{equation}

\subsubsection{Pre-quanta}
\label{subsubsec:preQuanta}
The pre-quanta stage encapsulates the conversion from an energy deposit to a number of produced photons and electrons. The model functions determining the probability distribution parameters are obtained from v2.2.2 of the NEST code \cite{szydagis_m_2021_5080263}, and we direct the reader to the references therein for further details of the physics. Here we summarise the probability distributions used in each block, and will direct the reader to Appendix \ref{appendixa} for detailed model descriptions. For these models, we assume a cylindrical TPC with a fixed fiducial volume of liquid, and only consider ER and NR events within the volume. 

Incorporation of these NEST yield models into the Flamedisx framework was not possible with a simple modification of the existing blocks coupled with a linear extension to additional blocks, as for the post-quanta models detailed in Section \ref{subsubsec:postQuanta}. Instead, two substantial modifications were made to the block performing this computation, shown in green in Figure \ref{fig:newblockstructure}. Firstly, its dimensionality was increased by one internally contracted dimension, capturing the splitting into ions and excitons before recombination occurs. Secondly, a number of these tensors are summed together over a set of relevant energies for each event, reflecting the parameterisation of NEST's yield models by `true' energy deposition. This is in contrast to the original Flamedisx models, where the yields are parameterised in terms of some pre-computed number of net electrons and photons produced. Both of these complications introduce memory usage and performance challenges, discussed further in Section \ref{subsec:performance}.

Let us consider the pre-quanta model block for the ER case. A normal distribution is used to model the fluctuations on the mean yields, producing $n^{\text{q}}_{\text{prod}}$ total quanta. From this, a binomial process models a number of produced ions $n^{\text{i}}_{\text{prod}}$. Finally, a skew normal distribution models the recombination fluctuations leading to $n^{\text{el}}_{\text{prod}}$, such that we can then obtain $n^{\text{ph}}_{\text{prod}}$ by subtracting $ n^{\text{el}}_{\text{prod}}$ from  as $n^{\text{q}}_{\text{prod}}$. Both the normal and skew normal distributions have continuity corrections accounted for. When dealing with the skew normal distribution, we need to account for the additional constraint the NEST models impose, that $n^{\text{el}}_{\text{prod}} \leq n^{\text{i}}_{\text{prod}}$. This is done at the level of the distribution, and is detailed fully in Appendix \ref{appendixb}.

In the NR case, a normal distribution models the production of $n^{\text{i}}_{\text{prod}}$ ions based on the mean yield, with a further normal distribution modelling the difference between the produced number of total quanta $n^{\text{q}}_{\text{prod}}$ and the value of $n^{\text{i}}_{\text{prod}}$. We can now obtain $n^{\text{el}}_{\text{prod}}$ which is modelled identically to the ER case, with just the forms of the model functions determining the parameters being different. Continuity corrections are applied here for all three distributions.

We construct the green tensor in Figure \ref{fig:newblockstructure} over suitable hidden variable values of the 3 dimensions $(n^{\text{el}}_{\text{prod}}, n^{\text{ph}}_{\text{prod}}, n^{\text{i}}_{\text{prod}})$. A fourth dimension is included if events are grouped into batches. This tensor is constructed for a specific value of the energy, $E$. Each element is then the product of the 3 probability elements: $P(n^{\text{q}}_{\text{prod}}|E)$, $ P(n^{\text{i}}_{\text{prod}}|E)$, and $ P(n^{\text{el}}_{\text{prod}}|E)$ for either ER or NR sources, where we indicate the explicit dependence on energy but not the other conditional dependencies seen in Figure \ref{fig:newblockstructure}, which are different for ER and NR. Energy dependence enters at the level of the mean electron, photon, exciton and ion yields, which are used in calculating distribution parameters, outlined more clearly in Appendix \ref{appendix:prequanta}.

Contracting each of these tensors internally over the $n^{\text{i}}_{\text{prod}}$ dimension results in a tensor over  $(n^{\text{el}}_{\text{prod}}, n^{\text{ph}}_{\text{prod}})$ which is constructed of probability elements $P(n^{\text{el}}_{\text{prod}}, n^{\text{ph}}_{\text{prod}} | E)$, defined as the probability of a certain ER or NR energy deposit to produce $n^{\text{el}}_{\text{prod}}$ electrons and $n^{\text{ph}}_{\text{prod}}$ photons, given the energy $E$ of the deposit. For each event, we multiply this at each energy by the value of the interaction rate spectrum of the $j^{th}$ source, $R^j(E)$, which may also be a function of event position and time for certain sources. We henceforth refer to this quantity as the energy spectrum. We then multiply this with the post-quanta blocks and repeat over $R^j(E)$. By summing these results together, we obtain $R^j(S1,S2,x,y,z,t)$. This can be repeated for all events, and all relevant signal/background sources, to allow for computation of the likelihood of the dataset. More detail on this is given in \cite{Aalbers_2020}.

\subsubsection{Post-quanta}
\label{subsubsec:postQuanta}

The post-quanta stage encapsulates the detection of the produced electrons/photons, as described in Section \ref{sec:TPCs}. We currently seek only to emulate NEST's `parametric' S1 calculation mode, where a detection threshold is not applied to individual PMT hits; rather, the DPE effect and a parametric detection efficiency is applied to the sum of detected photons. This leads to a marginally less accurate calculation at very low S1 signal sizes. We intend to incorporate the full calculation in a future version of FlameNEST, though encapsulating it within the tensor framework is not straightforward.

The first block in the lower row of the post-quanta blocks in Figure \ref{fig:newblockstructure} represents the binomial process which describes the number of photons detected, $n_{\text{det}}^{\text{ph}}$, given the number of photons produced, $n_{\text{prod}}^{\text{ph}}$, with a position-dependent detection probability. Detector threshold effects are also applied at this stage by introducing a minimum photon cut. It should be noted that the minimum photon cut is a requirement on the total number of detected photons, not accounting for the expected distribution of photons across PMTs, a feature modelled more fully by NEST and used in many experimental analyses. This will be implemented in future FlameNEST versions. The next block describes the binomial process by which the DPE effect may lead to a single detected photon producing two photoelectrons. The total number of photoelectrons is denoted $n_{\text{prod}}^{\text{phel}}$. This is followed by a binomial process which links $n_{\text{prod}}^{\text{phel}}$ to a number of detected S1 photoelectrons, $n_{\text{prod}}^{\text{S1-phel}}$. Finally, we apply a Gaussian smearing to $n_{\text{prod}}^{\text{S1 phel}}$ to obtain S1. Acceptance cuts can then be applied to the final S1 signal.

The first block in the upper row of the post-quanta blocks in Figure \ref{fig:newblockstructure} represents the binomial electron survival process during drift, whereby an electron may be lost due to interactions with impurities in the LXe. The number of electrons extracted to the gas region from the $n_{\text{prod}}^{\text{el}}$ produced electrons in the liquid region is denoted $n_{\text{det}}^{\text{el}}$. As previously discussed, these extracted electrons produce electroluminescence in the xenon gas. The number of photons produced from this process is denoted $n_{\text{prod}}^{\text{S2-ph}}$, with the process being described by a normal distribution with a continuity correction applied. We use another binomial, again with position-dependent detection efficiencies, to model the detection of a number $n_{\text{det}}^{\text{S2-ph}}$ of these photons. We introduce the DPE effect identically to the S1 case, leading to $n^{\text{S2-phel}}$ photoelectrons. A Gaussian smearing is applied to model the final S2 signal, before acceptance cuts can be applied.

\subsection{Performance features}\label{subsec:performance}
The modifications made to Flamedisx to fully capture the NEST models introduced a substantial speed penalty to the computation, necessitating the implementation of a number of additional features to mitigate this. This Section details these performance features.

\subsubsection{Generalising bounds computations}
\label{subsubsec:bounds}
As discussed in Section \ref{subsec:blockStructure}, for each data event Flamedisx must compute bounds on each hidden variable, determining the size of the tensors constructed. These must be large enough that all probability elements contributing non-negligibly to the sum in Equation \ref{eq:fullSum} are included, but not so large as to be redundantly including elements contributing close to 0. Flamedisx's original implementation of this needed improvement for two reasons: firstly, the calculations did not fully account for fluctuations in all distributions, and so the bounds had to be made particularly wide to ensure that full range of relevance of each hidden variable was captured; secondly, the calculation to produce the bounds needed to be reproduced each time a new model block was added, which in the case of some of the additional blocks added for FlameNEST was non-trivial.

We generalised the bounds computation procedure in Flamedisx to calculate the bounds for each block's input hidden variable, $I$, based on already calculated bounds for each block's output hidden variable, $O$. Bayes' theorem states

\begin{equation}
\label{eq:bayes}
    P(I=i | O=o) = \frac{P(O=o | I=i) P(I=i)}{P(O=o)},
\end{equation}

\noindent where the probability $P(O=o | I=i)$ is evaluated across the support of the input hidden variable, or some sensible restriction of this domain, for the already calculated bound values of the output hidden variable; that is, to calculate the lower bound on $I$, the lower bound of $O$ would be used, taking the converse for the upper bound. The prior probability $P(I=i)$ is by default flat, but certain blocks can override this when it improves the bound calculation procedure to do so. The prior is estimated via drawing values of the hidden variable $I$ from a large pre-computed Monte Carlo reservoir, filtering as appropriate based on already computed bounds. An example of this for the FlameNEST block structure is given shortly.

\begin{figure}[H]
    \centering
    \includegraphics[width=14cm]{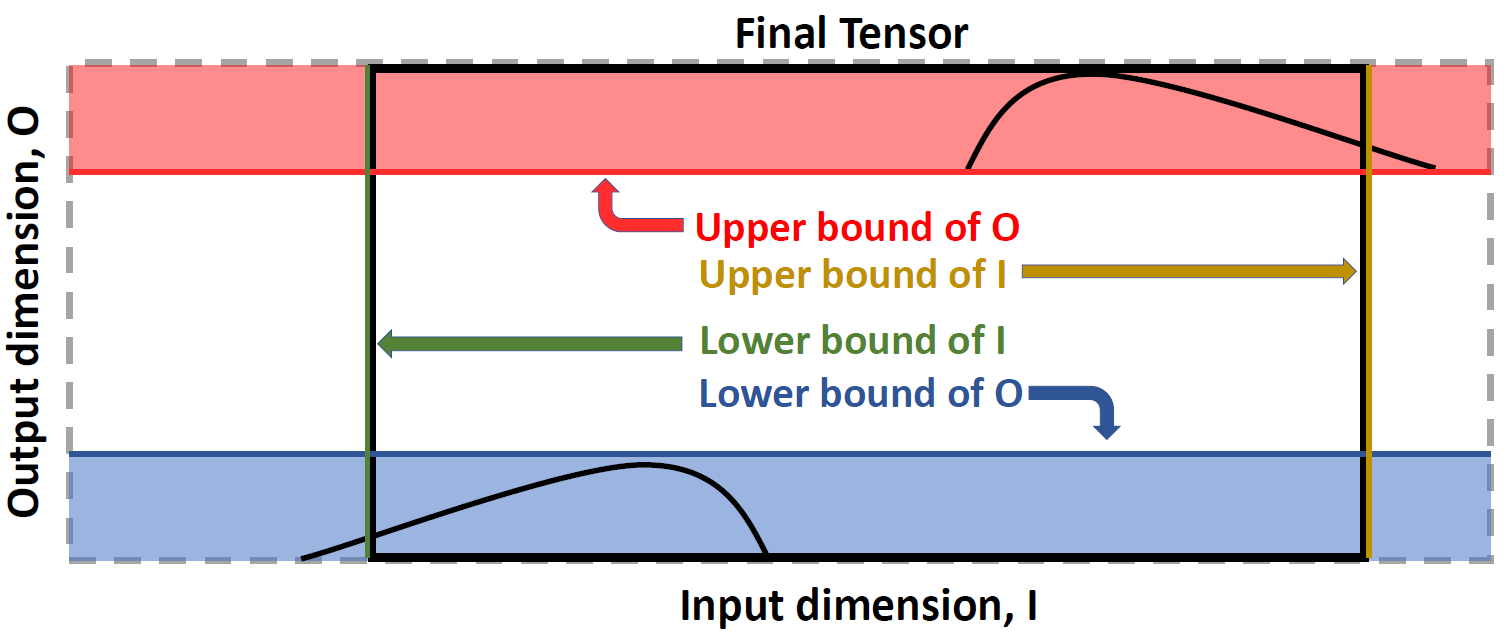}
    \caption{Pictorial demonstration of the bounds computation for a block. The lower and upper bounds on the output dimension, $O$, are used to determine the input distributions, $P(I=i, O=o_{\text{max}}$) and $P(I=i, O=o_{\text{min}}$), respectively, represented here as the black curves. We can determine the lower and upper bounds on the input dimension using these distributions depending on the max sigma chosen by the user. The final tensor is shown as a black box.}
    \label{fig:boundsimage}
\end{figure}

Bounds on $I$ can then be obtained by constructing the cumulative distribution function of the posterior probability $f(x)$ over the support of $I$, $x \in \{supp(f)_{min}, supp(f)_{max}\}$,

\begin{equation}
    F(x) = \frac{1}{\mathcal{N}} \sum_{i=supp(f)_{min}}^{x} f(i),
\end{equation}

\noindent with an appropriate normalisation factor $\mathcal{N}$ chosen such that $f(x)$ is normalised to 1 and we can set the denominator in Equation \ref{eq:bayes} to unity. The lower and upper bounds are then taken as the values of $x$ where $F(x)$ evaluates to some user-defined low and high values of probability, where taking more extreme values corresponds to calculating wider bounds. This is depicted pictorially in Figure \ref{fig:boundsimage}.

The method proceeds by computing the bounds for each block recursively -- bounds on the outermost hidden variables are computed based on the observables, then the procedure outlined is repeated for each preceding block in turn until bounds are computed on all hidden variables. In the case of the FlameNEST block structure, we make two modifications to the above procedure, made to improve the accuracy of the tensor and energy stepping outlined in Sections \ref{subsubsec:tensorStepping} and \ref{subsubsec:energyStepping}.

The first is making a manual calculation of the ion bounds. As we construct the central quanta tensor for various values of the energy, contracting over the ion dimension for each before summing them together, it is possible to choose the ion bounds to be different for each summed energy. Therefore the ion bounds are estimated directly as a function of energy for each summed tensor, as outlined in Appendix \ref{appendixc}. Whilst in principle the Bayesian procedure could be used instead, it was found that a manual calculation in this case substantially improved performance, being of reliable accuracy due to the proximity of this hidden variable to the input dimension, energy.

The second change is that an additional bounds estimation is made for the energy values to be summed over when constructing the central quanta tensor. This is done by filtering the same MC reservoir used to calculate the priors within the bounds of electrons and photons produced, for each event, and taking (user-defined) extremal quantiles of the resulting distribution of energies to estimate bounds on the energy.

One can summarise the bounds computation for the FlameNEST block structure as follows. We use the Bayesian inversion procedure to calculate bounds for all hidden variables in the post-quanta blocks of Figure \ref{fig:newblockstructure}, taking flat priors in each case. We then compute preliminary bounds on electrons and photons produced using the same procedure, taking a flat prior. Once these have been obtained, energy bounds can be obtained for each event using the procedure detailed above. These energy bounds are then used together with the bounds on the outermost hidden variables - S1 and S2 photoelectrons detected - to obtain priors on electrons and photons produced. These are then used to obtain a second, tighter set of bounds on electrons and photons produced. Finally, ion bounds are computed using the procedure outlined in Appendix \ref{appendixc}.

\subsubsection{Variable tensor stepping}
\label{subsubsec:tensorStepping}
Originally, Flamedisx would construct each hidden variable dimension in unit steps between the computed bounds. This size of the tensors for high energy events, even for the original Flamedisx models, would thus become too large to fit in memory on many GPUs. For FlameNEST, the introduction of a number of additional post-quanta model blocks, as well as the pre-quanta block with an internally contracted dimension, greatly compounded this problem. In order to allow \texttt{TensorFlow} to hold all the tensors for the computation in memory and to speed up the Flamedisx computation, we implemented a variable stepping over the hidden variables.

A maximum dimension size may be specified for any set of hidden variables, and if the difference between the upper and lower bounds for any events is greater than this, the tensors constructed for that event batch will have hidden variable dimensions increasing in integer steps greater than 1. These steps are chosen such that no hidden variable dimension goes above its maximum dimension size. Provided that all distributions computed over a stepped hidden variable are sufficiently smoothly varying over the stepped values, each calculated probability element may simply be re-scaled by the step size of its domain, with the overall computation then returning a result approximately the same as if no stepping had been done.

\subsubsection{Variable energy stepping}
\label{subsubsec:energyStepping}
As detailed in Sections \ref{subsec:blockStructure} and \ref{subsubsec:bounds}, the green quanta tensor in Figure \ref{fig:newblockstructure} is constructed across energies between the energy bounds for each source/event pair. Provided the energy bounds are chosen to be wide enough, terms outside of the bounds will contribute negligibly to the sum over $E$ in Equation \ref{eq:fullSum}. 

To further accelerate the computation, provided that the shape of the source's energy spectrum is smoothly varying within these bounds, it is possible to obtain an accurate value of $R^j(S1,S2,x,y,z,t)$ by taking larger steps in $E$ in the sum, re-weighting each $R^j(E)$ by the step size taken relative to the energy granularity of the spectrum. This is analogous to the variable tensor stepping described in Section \ref{subsubsec:tensorStepping}.

\subsubsection{Model-dependent approximations}

As discussed in Sections \ref{subsubsec:preQuanta} and \ref{subsubsec:postQuanta}, it is necessary to apply continuity corrections and account for the constraint that $n^{\text{el}}_{\text{prod}} \leq n^{\text{i}}_{\text{prod}}$ to ensure good matching between the FlameNEST model implementation and the NEST MC models. However, above certain energy thresholds this becomes redundant, and has little effect on the accuracy of the computation. Therefore, both of these aspects are ignored when calculating quanta tensors above 5 keV for ER sources and 20 keV for NR sources. For dramatically different detector conditions, the user may wish to verify that these thresholds remain sensible choices.

\section{Validations}
For the performance features outlined in Section \ref{subsec:performance} to be used in practise, in must first be verified that they still produce accurate computed values of $R^j(S1,S2,x,y,z,t)$ for all sources $\{j\}$ of interest at a range of energies, whilst providing ample speedup to the computation. This Section presents the results of a series of such validations.

\subsection{Mono-energetic sources}
\label{subsec:mono_validations}

In order to validate the FlameNEST computation, we compare directly with the result from a finely binned, high statistics NEST v2.2.2 simulation at 1, 10, and 100 keV energies. Whilst the approximations outlined in Section \ref{subsec:performance} will introduce some error in the calculation compared to the idealised case of infinite bounds and no stepping, if the difference between the FlameNEST result and a Monte Carlo template-estimated differential rate is sufficiently small, this can be accepted. The reason for this is twofold; firstly, parameters in the NEST models come with, in some cases, very large errors, and shifts in the differential rate coming from approximations in the FlameNEST computation can be absorbed by small shifts in these parameters. Secondly, MC templates come with their own errors: errors from finite simulation statistics, binning and template interpolation as nuisance parameters are floated, meaning small errors in likelihood evaluation are not unique to FlameNEST.

We start by filling a two-dimensional histogram in (S1,S2) at a fixed event position and time, to avoid the computational cost of achieving sufficent simulation statistics with a 6-dimensional template, a reminder of why the Flamedisx computation is superior to a template computation. We take the NEST defaults for all parameters, which is the LUX detector's third science run \cite{LUXRun3}, and fix all sources at the centre of this detector. The histogram is filled with $1\times10^{8}$ NEST events with 50 logarithmically-spaced bins in both dimensions.

In order to calculate a differential rate from this histogram, we divide the number of events in each bin by the bin volumes and the total number of MC events in the template, and multiply it by the total number of expected events using some arbitrary exposure, after all data selection cuts are applied. We evaluate the FlameNEST differential rate at the centre of each bin and at the fixed position and time, and for each bin plot the difference between the FlameNEST differential rate and the MC template differential rate, normalised by the estimated error from the MC template calculation. This includes an estimation of the (Poisson) error from finite simulation statistics in each bin, assuming bins are uncorrelated, and an estimation of the binning error, obtained by also calculating the FlameNEST differential rate at the corner of each bin. For all subsequent validations we take 3$\sigma$ bounds, such that the Bayesian bounds procedure uses probability corresponding to the 3$\sigma$ quantile of a Gaussian distribution, and choose all tensors to have a maximum dimension size of 70.

Figure \ref{fig:mono_validations} shows the comparison described above for mono-energetic ER and NR sources, respectively. Both ER and NR sources at all energies show a good agreement. Any small offsets or shape to the distributions are a result of the finite tensor bounds and the tensor stepping outlined in section \ref{subsec:performance}, however they are well-within the errors inherent to template-based likelihood evaluation. 

We recommend this validation process is repeated when further model changes are implemented in FlameNEST. Smaller changes to models might not carry the same significance at all energies so we also recommend a wide scan in energy space.

\begin{figure}[H]
    \centering
    \begin{minipage}{0.45\textwidth}
        \centering
        \includegraphics[width=1.1\textwidth]{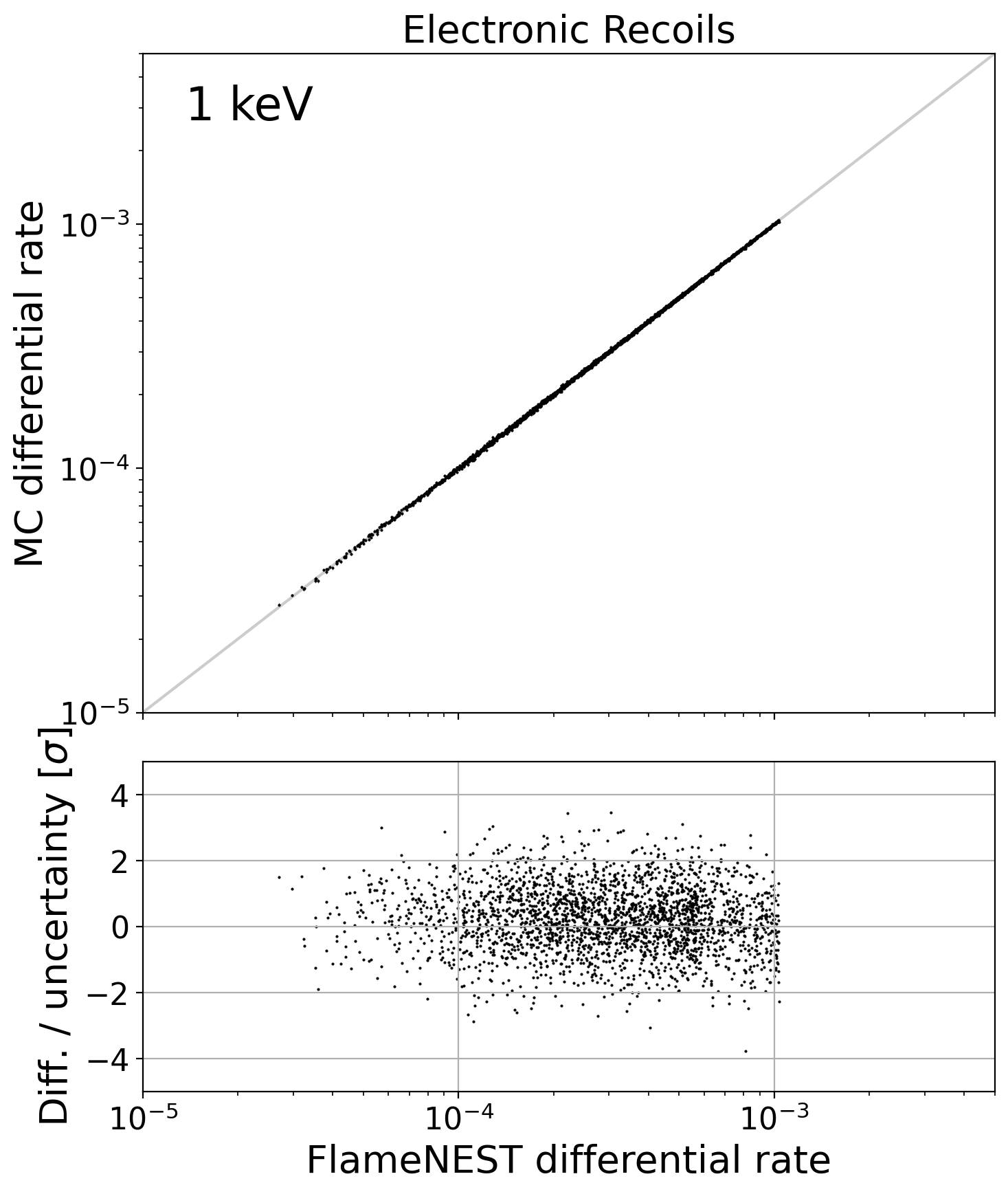}
    \end{minipage}\hfill
    \begin{minipage}{0.45\textwidth}
        \centering
        \includegraphics[width=1.1\textwidth]{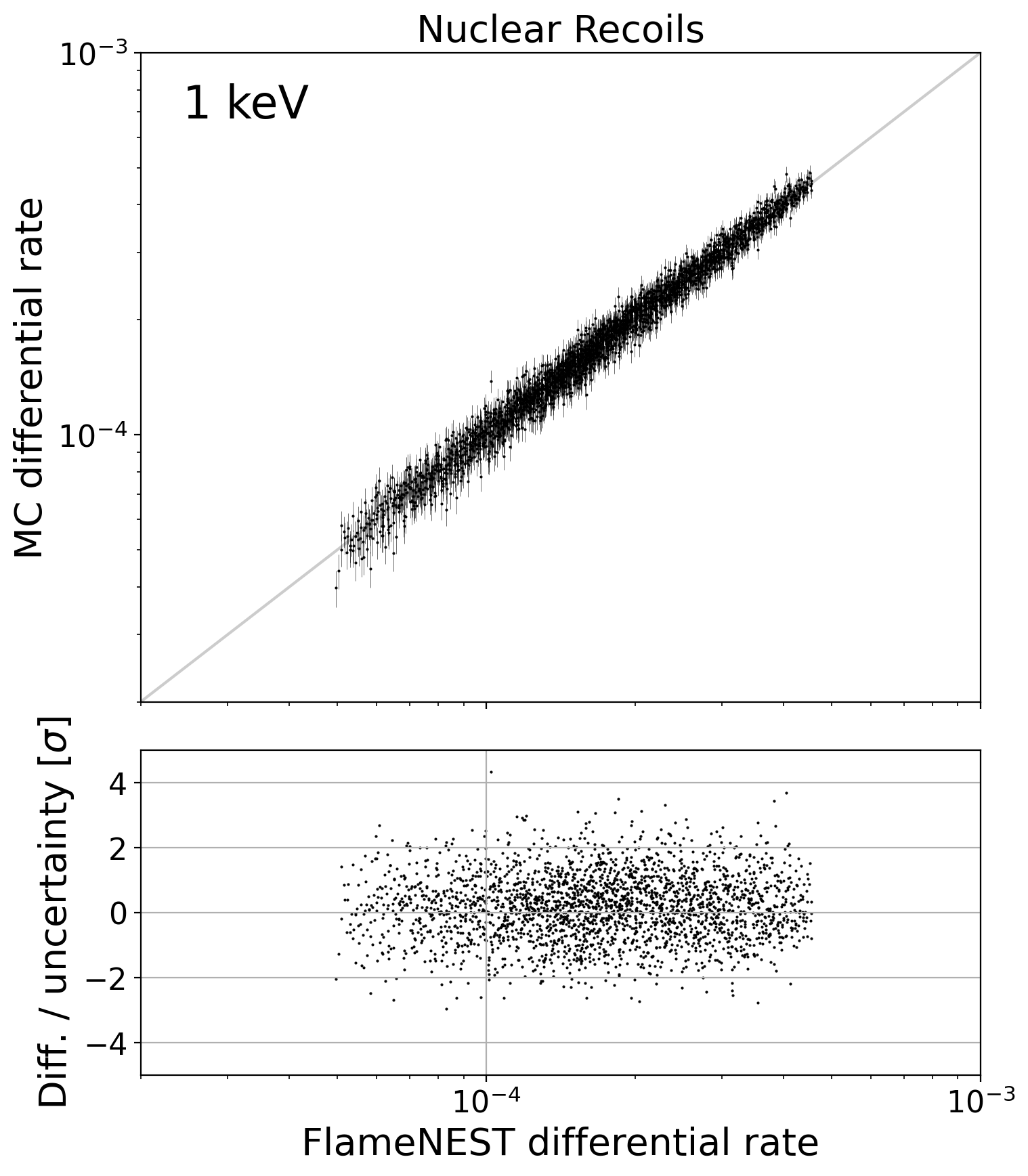}
    \end{minipage}
    \begin{minipage}{0.45\textwidth}
        \centering
        \includegraphics[width=1.1\textwidth]{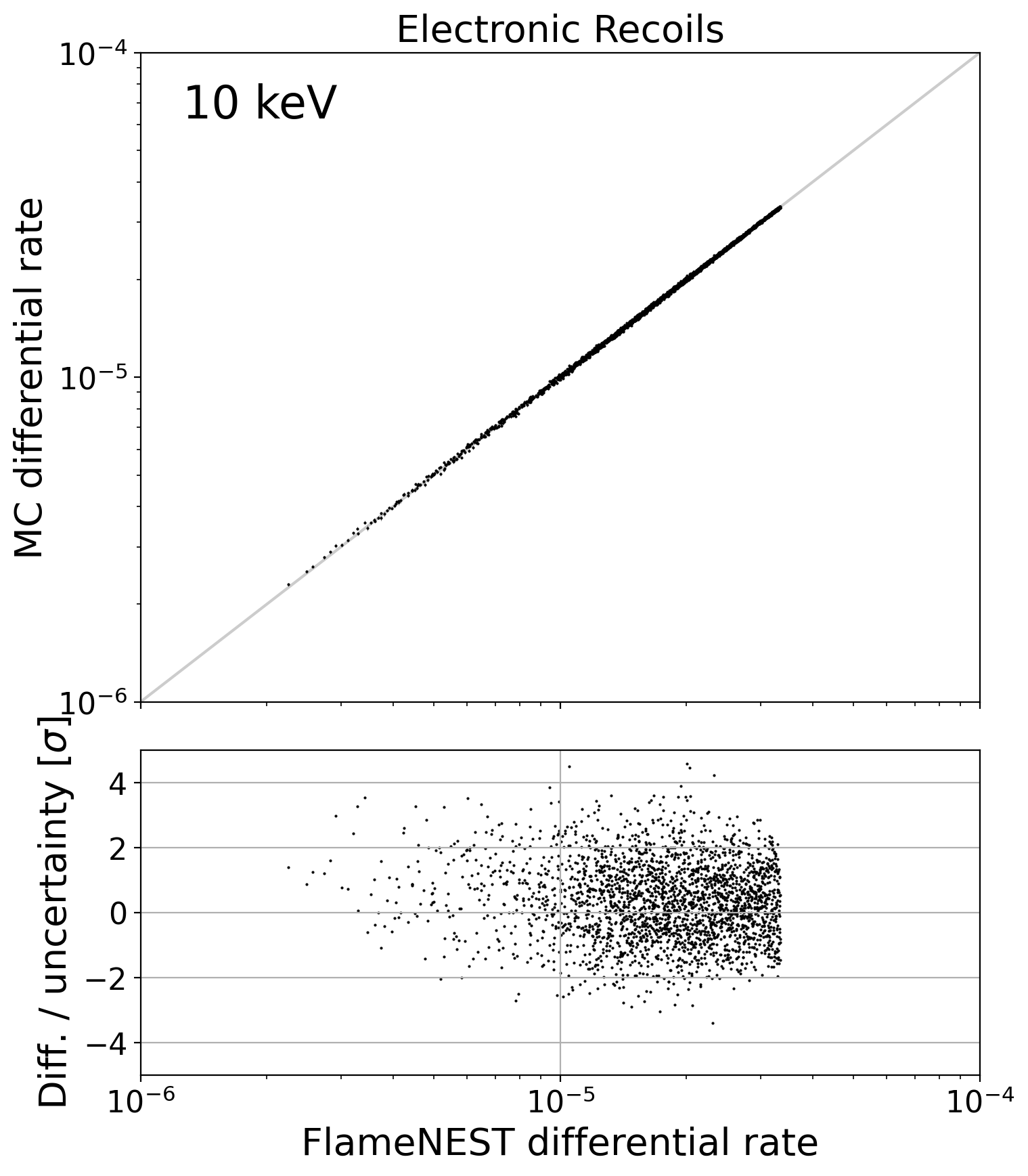}
    \end{minipage}\hfill
    \begin{minipage}{0.45\textwidth}
        \centering
        \includegraphics[width=1.1\textwidth]{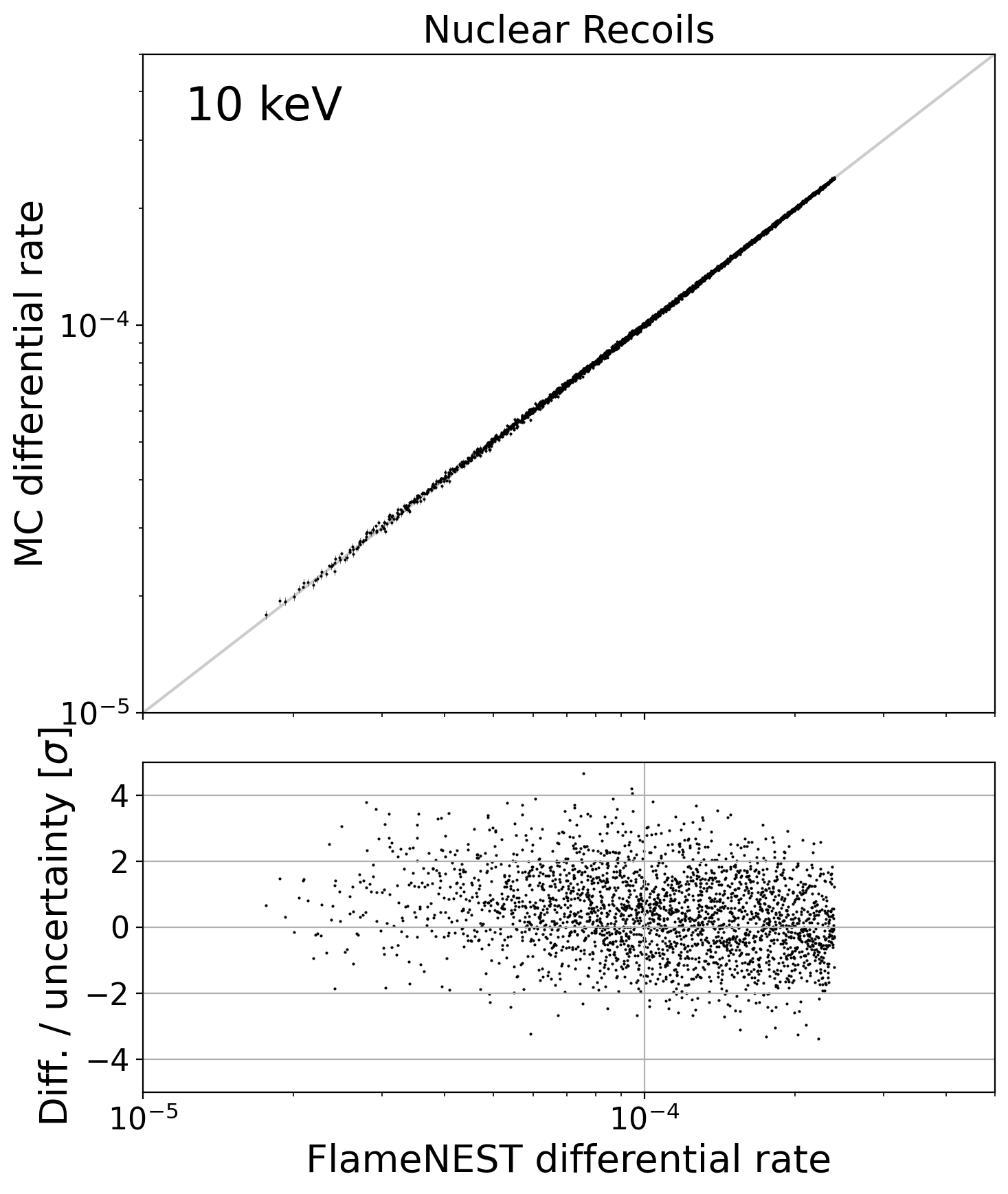}
    \end{minipage}
\end{figure}
\begin{figure}[H]\ContinuedFloat
    \begin{minipage}{0.45\textwidth}
        \centering
        \includegraphics[width=1.1\textwidth]{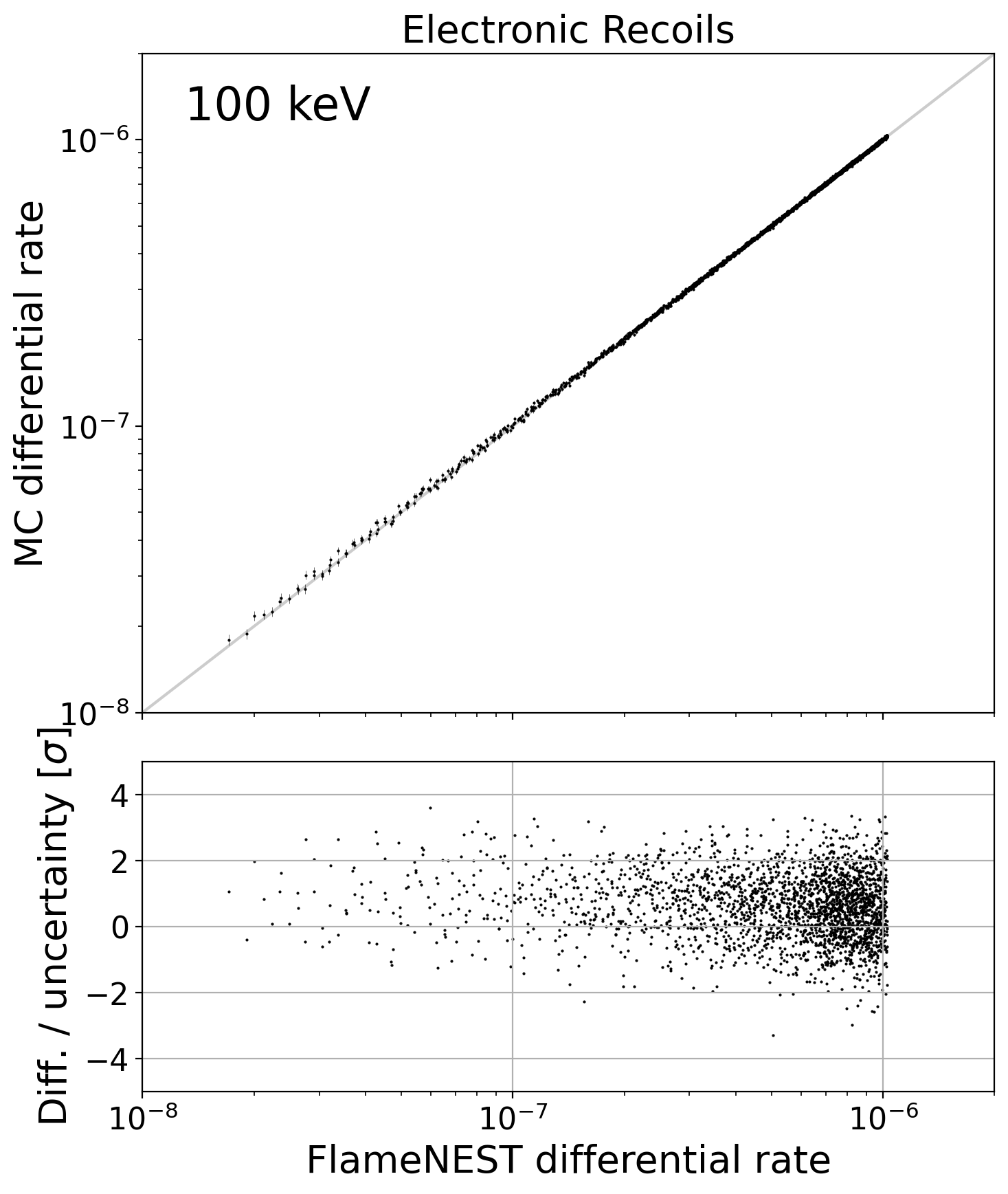}
    \end{minipage}\hfill
    \begin{minipage}{0.45\textwidth}
        \centering
        \includegraphics[width=1.1\textwidth]{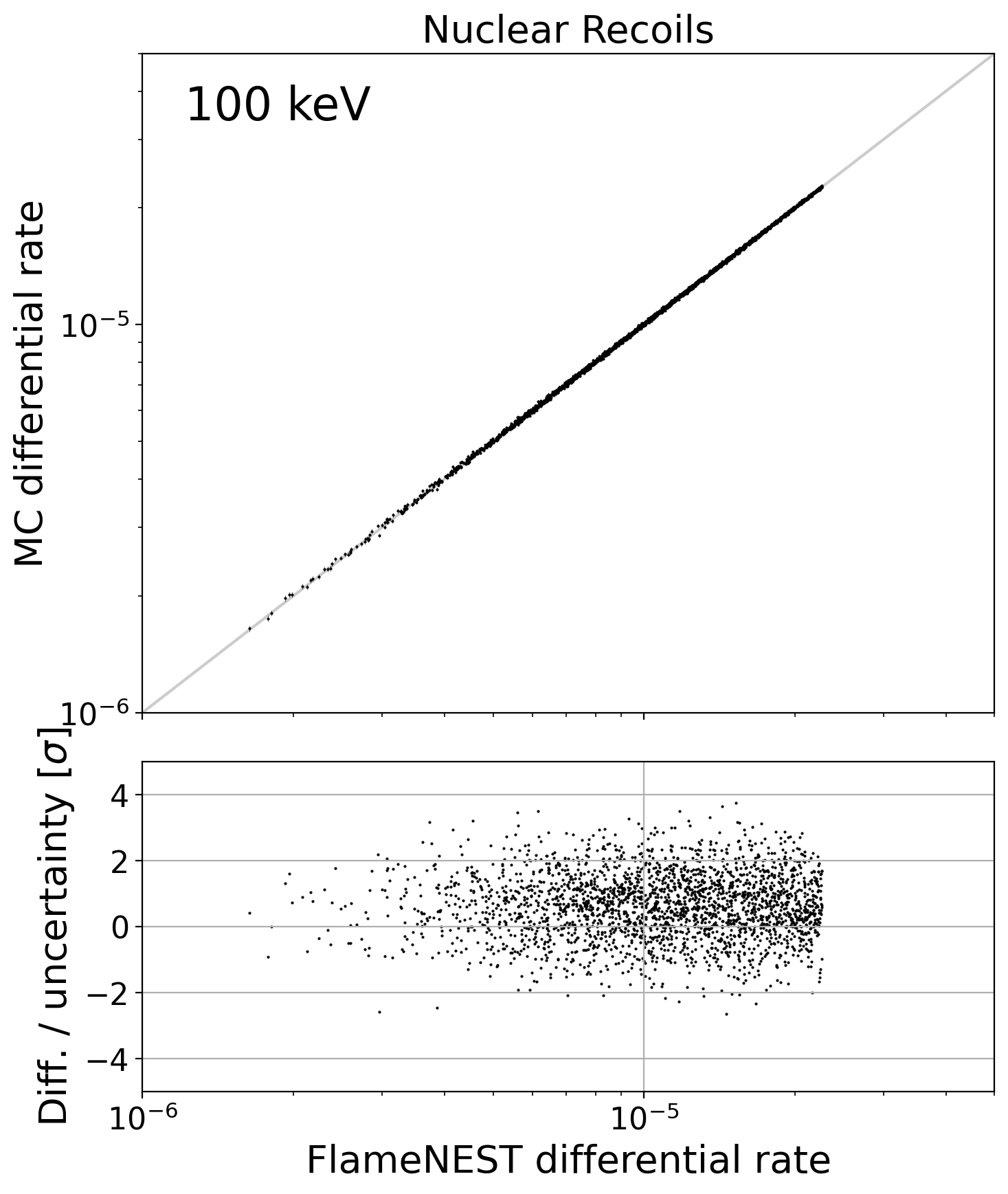}
    \end{minipage}
    \caption{Difference between the FlameNEST and MC template differential rate for bins in S1/S2 space for 1, 10 and 100 keV mono-energetic ER and NR sources, fixed at the centre of the LUX detector, presented in terms of the estimated Poisson statistics + binning error from the MC template calculation.}\label{fig:mono_validations}
\end{figure}

\subsection{Full energy spectra}
\label{subsec:uniform_validations}

As described in Section \ref{subsubsec:energyStepping}, FlameNEST will step over the energies remaining once the input spectrum of the source has been trimmed between the calculated energy bounds for each event (or batch of events). Here, we demonstrate how this stepping impacts the speed and accuracy of the computation. For ER and NR sources we run the same computation as in Section \ref{subsec:mono_validations}, this time simulating a flat energy spectrum between 0 - 100 keV using NEST. When doing the FlameNEST computations we vary the maximum energy dimension size -- this caps the size of the trimmed spectrum between the energy bounds, applying a stepping if the size of the trimmed spectrum is above the specified maximum. We set the full flat spectra used in the FlameNEST computation to have 1000 sampling points initially. All other parameters are the same as described in Section \ref{subsec:mono_validations}, except for the maximum dimension size of the internally contracted ions produced dimension, which is now capped at 30. We found that the resulting speed increase justified the minimal loss in accuracy of the FlameNEST computation, especially when the effects on accuracy of the energy stepping are accounted for.

To quantify the overall accuracy at different maximum energy dimension sizes, we define an accuracy metric, $\Delta$, over the template bins to be a weighted average over all bins of the percentage difference in differential rate between the FlameNEST computation and the template evaluation, weighted by the averaged differential rate of that bin, as in Equation \ref{eq:accuracyMetric}. Here, $R(\text{S1},\text{S2})_{\text{FN/MC}}$ denotes the differential rate at the bin with centre $(\text{S1},\text{S2})$ using the FlameNEST / Monte Carlo template evaluation, and the sum is over all template bins. We chose this over the accuracy metric used in section \ref{subsec:mono_validations} to avoid added difficulties in estimating the Poisson error on the template bin values that occur when the templates become particularly large in S1/S2 space, arising from the fact that correlations across bins can become particularly strong. This choice of metric also avoids the issue of most bins being empty for templates covering the full observable space when using such a broad energy spectrum.

\begin{equation}
\label{eq:accuracyMetric}
    \Delta = \frac{\sum_{\text{S1},\text{S2}}(R(\text{S1},\text{S2})_{\text{MC}} - R(\text{S1},\text{S2})_{\text{FN}})}{\sum_{\text{S1},\text{S2}}\frac{1}{2}(R(\text{S1},\text{S2})_{\text{MC}} + R(\text{S1},\text{S2})_{\text{FN}})} \times 100\%
\end{equation}

\begin{figure}
    \begin{minipage}{.5\textwidth}
        \centering
        \includegraphics[width=1.\linewidth]{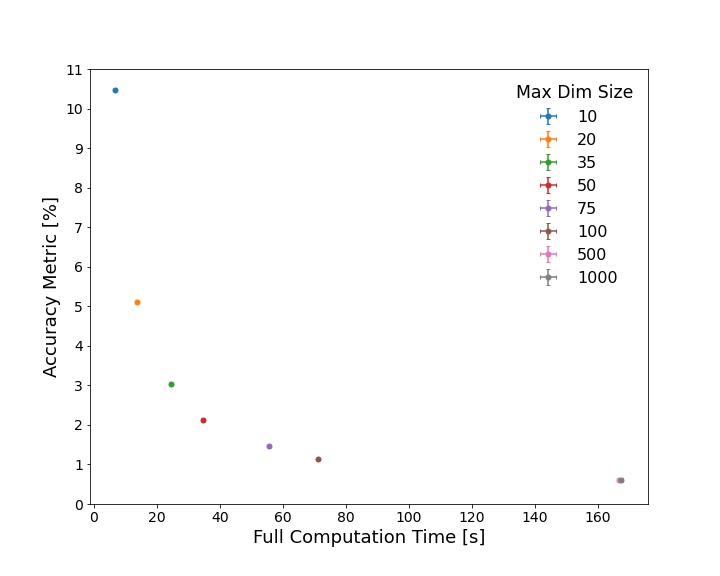}
    \end{minipage}
    \begin{minipage}{.5\textwidth}
        \centering
        \includegraphics[width=1.\linewidth]{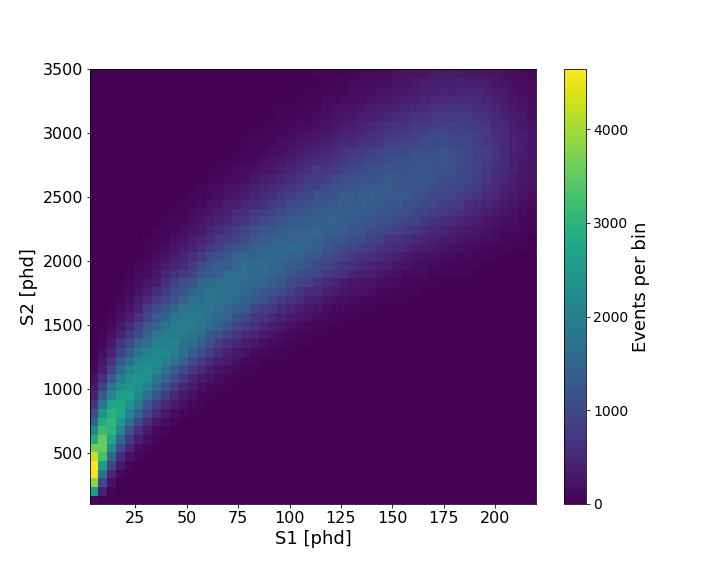}
    \end{minipage}
    \caption{Accuracy metric vs full computation time for a range of different maximum energy dimension sizes for an NR source with a flat energy spectrum between 0.01 and 100 keV, using LUX detector parameters and fixed at the centre of this detector. The resulting (S1,S2) template used for one of the 10 comparisons is also shown. Approximately 1750 bins are used for the computation after the empty bins are removed.}\label{fig:nr_validations}
\end{figure}

\begin{figure}
    \begin{minipage}{.5\textwidth}
        \centering
        \includegraphics[width=1.\linewidth]{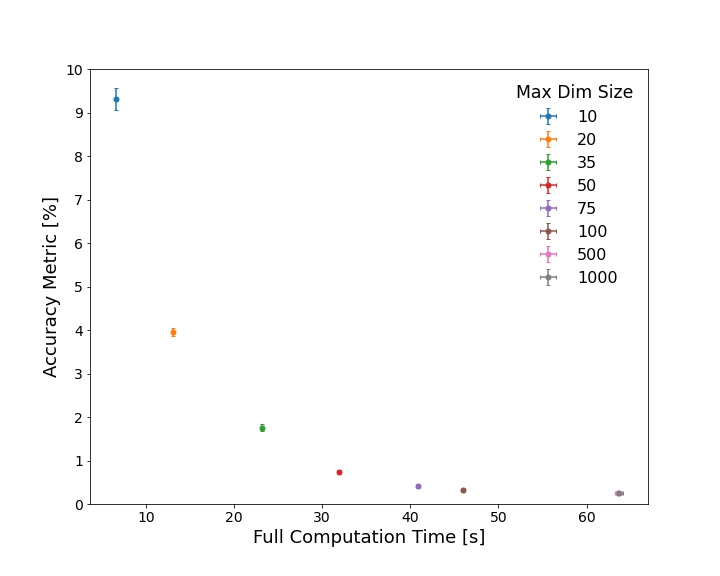}
    \end{minipage}
    \begin{minipage}{.5\textwidth}
        \centering
        \includegraphics[width=1.\linewidth]{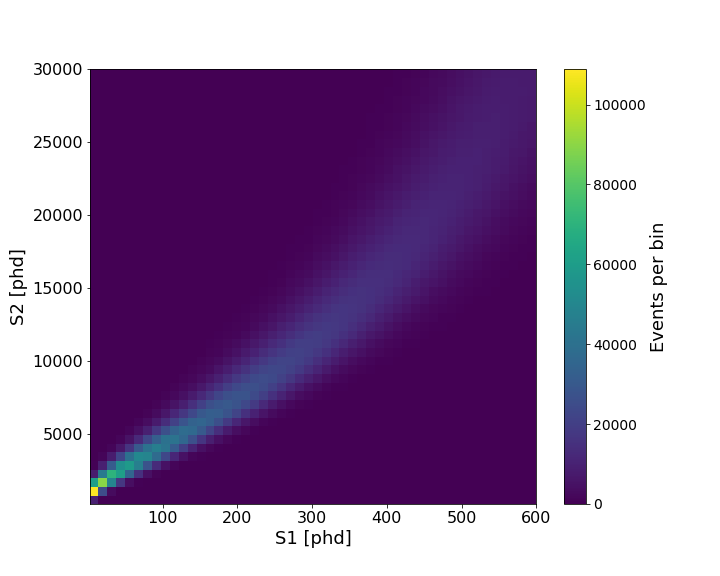}
    \end{minipage}
    \caption{Accuracy metric vs full computation time for a range of different maximum energy dimension sizes for an ER source with a flat energy spectrum between 0.01 and 100 keV, using LUX detector parameters and fixed at the centre of this detector. The resulting (S1,S2) template used for one of the 10 comparisons is also shown. Approximately 1000 bins are used for the computation after the empty bins are removed.}\label{fig:er_validations}
\end{figure}

Figures \ref{fig:nr_validations} and \ref{fig:er_validations} present the resulting accuracy metric value for each energy maximum dimension size, plotted against the computation time to evaluate the FlameNEST differential rate across bins for the ER and NR spectra shown. The computation is repeated for 10 separate NEST templates to estimate the variation seen. Bins with 0 MC template events are discarded from the computation; after doing so, approximately 1000 bins remained for the ER source and approximately 1750 bins remained for the NR source. We benchmark using a Tesla P100 GPU.

Unsurprisingly the computation time increases as more energy steps are added, though perfect linearity is not seen as the number of events (bins) per computational batch is altered each time to maximise usage of the GPU memory. The accuracy metric behaves as expected; it is up to the user to decide the desired degree of accuracy, and to pay the corresponding cost in computation time.

Saturation in time and accuracy is ultimately seen above a maximum energy dimension size; this happens when (for the majority of bins) the size of the input spectrum within the energy bounds is smaller than this maximum dimension size, rendering energy stepping redundant here. At this stage the remaining discrepancy in differential rate comes down to the other approximations made; the tensor stepping, the hidden variable and energy bounds computations and the number of terms used in the expansion of Owen's T function, the calculation of which is necessary for the FlameNEST models (see Appendix \ref{appendixb}).

The calculated accuracy metric will differ for energy spectra with more features; here, the user would likely want to implement a variable maximum energy dimension size, taking it to be larger for events where the energy bounds cover regions of the spectrum with more features. Performing this same test would then allow them to validate that they are achieving sufficient accuracy for their source spectra.

\begin{figure}
    \begin{minipage}{.5\textwidth}
        \centering
        \includegraphics[width=1.\linewidth]{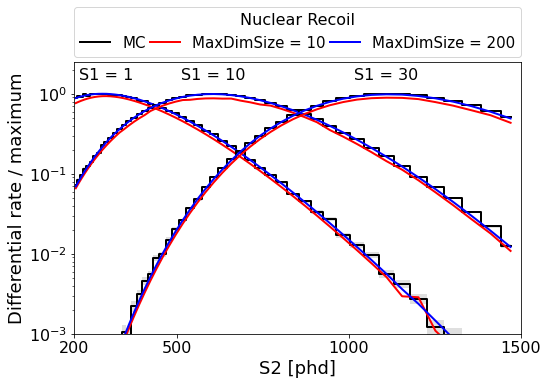}
    \end{minipage}
    \begin{minipage}{.5\textwidth}
        \centering
        \includegraphics[width=1.\linewidth]{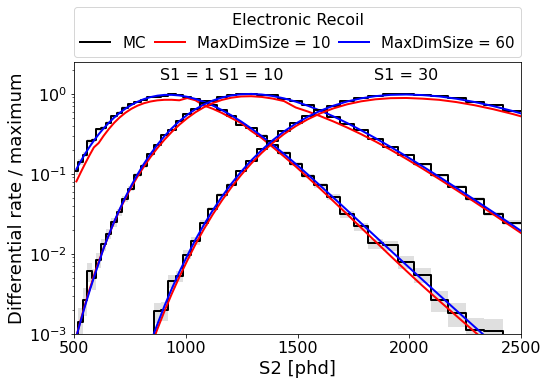}
    \end{minipage}
    \caption{MC and FlameNEST differential rates over S2 bins of 3 different S1 slices of the templates shown in Figures \ref{fig:nr_validations} and \ref{fig:er_validations}. We calculate the FlameNEST differential rates at two different maximum energy dimension sizes, to show the effect of this. We also depict for each bin the estimated Poisson statistics + binning error from the MC template calculation.}\label{fig:s1slices}
\end{figure}

To verify that presenting our results in terms of a weighted accuracy metric does not mask potential discrepancies at the tails of the distributions, we show in Figure \ref{fig:s1slices} the MC differential rate over S2 bins of 3 different S1 slices in each template, depicting also for each bin the estimated Poisson statistics + binning error from the MC template calculation. We overlay the FlameNEST differential rates at two different maximum energy dimension sizes; a poor choice for each as well as the choice for each that takes the corresponding accuracy metric value below 1\%. As can be seen, for the higher maximum dimension sizes, no discrepancies can be seen outside of the MC errors, whereas for the low maximum dimension size (and thus greater sized energy spectrum steps), more significant disagreement is observed.

Finally, we wish to provide an indication of how the performance of FlameNEST compares with the original benchmarking of Flamedisx presented in \cite{Aalbers_2020}. For a 0-10 keV ER source using a Tesla P100 GPU, we measure a differential rate computation time of 30ms per event, using a choice of 50 for the maximum energy dimension size following our findings in Figure \ref{fig:er_validations}. This is roughly a factor of 100 slower than the original models, whose benchmarking was additionally performed on a less modern GPU. In spite of this, it is important to reiterate that likelihood evaluation with 6 observables and multiple nuisance parameters is simply unfeasible using template methods, as the generation timescales become geological in magnitude. The vastly improved accuracy and applicability of the NEST models in the Flamedisx framework enables such computations to be performed confidently in a range of experiments, even if extra computation power must be sought to do so. It should also be noted that for the time-consuming step of test statistic estimation, asymptotic estimation methods can be appealed to, and further optimisations may be possible even in the case of doing the full MC toy estimation procedure, as long as the accuracy is carefully tracked.

\section{Conclusion}
We present FlameNEST, an amalgamation of Flamedisx and NEST. The technical challenges of this union and the subsequent performance has been described in detail. FlameNEST will allow for high-dimensional likelihood evaluation, increasing the physics reach of LXe dual phase TPC experiments. Furthermore, the incorporation of the NEST models will reduce the need for involved modifications of the models to fit real experimental data, as NEST models have been built to provide global fits to many existing datasets across multiple experimental setups.

Inter-collaborative analyses have in the past been difficult due to software differences and the ways different experiments handle their nuisance parameters. We believe FlameNEST will make future inter-collaborative efforts much simpler by providing a robust framework which can be straightforwardly adapted to each experiment. Such collaborative efforts will greatly facilitate the development of the next generation of noble element detection experiments, which in the case of LXe experiments will likely consist of a single, unified effort focused on one detector.

We point the reader to \url{https://github.com/FlamTeam/flamedisx}, where all of the FlameNEST code can be found within the original Flamedisx repository.

\acknowledgments

Funding for this work is supported by the U.K. Science and Technology Facilities Council under the contract numbers ST/S000844/1, ST/S505675/1, ST/S000666/1, and ST/S555360/1. We acknowledge additional support from the Cosmoparticle Initiative at University College London, the UCL Cities Partnership programme, Stockholm University and the Kavli Institute for Particle Astrophysics and Cosmology.

We would like to thank Matthew Szydagis and Gregory Rischbieter of the University of Albany for their guidance and advice regarding the Noble Element Simulation Technique. 

\newpage
\printbibliography

\newpage
\appendix
\section{Model Details} \label{appendixa}
Here we provide a detailed description of the distributions and parameters in the FlameNEST block structure.

\subsection{Model parameters}\label{appendix:params}
In this section, we will define the parameters which are used in the FlameNEST distributions. 

Table \ref{table:physical} lists the detector parameters which are typically measured or fixed and therefore unlikely to be floated as nuisance parameters in an analysis. It should be noted that the liquid electric field can in principle be position- and time-dependent.

\begin{table}[H]
\centering
    \caption{\label{table:physical}Physical, likely fixed, inputs to the FlameNEST model functions.}
    \begin{tabu}{| X[c] | X[c] |}
    \hline
    \textbf{Symbol} & \textbf{Meaning} \\ \hline
    $T$ & LXe temperature \\ \hline
    $P$ & LXe pressure \\ \hline
    $\epsilon_{\text{liq}}(x,y,z,t)$ & Liquid electric field \\ \hline
    $\epsilon_{\text{gas}}$ & Gas electric field \\ \hline
    $z_{\text{topDrift}}$ & Liquid/gas interface height \\ \hline
    $\Delta_{\text{gas}}$ & Distance between liquid/gas interface and anode \\  \hline
    $N_{\text{PMT}}$ & Number of PMTs \\ \hline
    \end{tabu}
\end{table}

NEST uses some of the parameters in Table \ref{table:physical} to calculate other fixed parameters used by the model functions. These are summarised in Table \ref{table:calculated}.

\begin{table}[H]
\centering
    \caption{\label{table:calculated}Calculated, likely fixed, quantities in the FlameNEST model functions.}
    \begin{tabu}{| X[c] | X[c] |}
    \hline
    \textbf{Symbol} & \textbf{Meaning} \\ \hline
    $\rho_{\text{liq}}(T,P)$ & Liquid xenon density \\ \hline
    $\rho_{\text{gas}}(T,P)$ & Gaseous xenon density \\ \hline
    $v_{\text{drift}}(\epsilon_{\text{liq}},\rho_{\text{liq}},T)$ & Electron drift velocity \\ \hline
    \end{tabu}
\end{table}

The post-quanta model functions take a number of parameters that will likely only be determined approximately in auxiliary measurements and thus should be floated as nuisance parameters in a statistical analysis. Table \ref{table:floated_postQuanta} lists these.

A `Fano factor' is used to account for an over-dispersion in S2 electroluminescence photons produced beyond Poisson statistics. The photon detection efficiencies determine the (binomial) detection probabilities for photons produced in liquid (S1) and gas (S2). Similarly the photoelectron detection efficiency determines the (binomial) detection probability for a single PMT to detect an (S1) photoelectron. The single photoelectron resolution coupled with the S1 and S2 noise terms determines the smearing of the final signals for a given number of detected photoelectrons due to PMT effects and electronics noise.

\begin{table}[H]
\centering
    \caption{\label{table:floated_postQuanta}Parameters that will likely be floated in the post-quanta FlameNEST model functions.}
    \begin{tabu}{| X[c] | X[c] |}
    \hline
    \textbf{Symbol} & \textbf{Meaning} \\ \hline
    $p_{\text{dpe}}$ & Double photoelectron emission probability \\ \hline
    $\tau$ & Electron lifetime \\ \hline
    $\mathcal{F}_{\text{S2}}$ & S2 Fano factor \\ \hline
    $g1$ & Photon detection efficiency in liquid at detector centre \\ \hline
    $g1_{\text{gas}}$ & Photon detection efficiency in gas \\ \hline
    $\mu_{\text{spe}}$ & Single photoelectron detection efficiency \\ \hline
    $\sigma_{\text{spe}}$ & Single photoelectron resolution \\ \hline
    $\Delta_{\text{S1}}$ & S1 noise \\ \hline
    $\Delta_{\text{S2}}$ & S2 noise \\ \hline
    \end{tabu}

\end{table}

Acceptance cuts are applied to the detected signals which may be accounted for in the models in the same way as the original Flamedisx structure. Parameters determining these are summarised in table \ref{table:selection}.

\begin{table}[H]
\centering
    \caption{\label{table:selection}Selection parameters.}
    \begin{tabu}{| X[c] | X[c] |}
    \hline
    \textbf{Symbol} & \textbf{Meaning} \\ \hline
    $S1_{\text{min}}$ & Minimum S1 acceptance \\ \hline
    $S1_{\text{max}}$ & Maximum S1 acceptance \\ \hline
    $S2_{\text{min}}$ & Minimum S2 acceptance \\ \hline
    $S2_{\text{max}}$ & Maximum S2 acceptance \\ \hline
    $\gamma_{\text{min}}$ & Minimum photons detected \\ \hline
    \end{tabu}

\end{table}

Table \ref{table:yieldParams} lists the parameters used by the model functions calculating the parameters of the yield probability distributions. They are all, directly or indirectly, functions of energy $E$, hence the need for the green tensor in Figure \ref{fig:newblockstructure} to be constructed for all relevant energies for an event and summed together.

Mean yields are calculated deterministically for both electrons and photons, along with the ratio of mean exciton yield to mean ion yield. The parameter $\alpha$, used as a distribution parameter for ER and NR, is defined as $\alpha = (1 + r_{\text{ex}})^{-1}$. The ER case calculates a `Fano factor' to model over-dispersion in quanta production beyond Poisson statistics. Finally a number of parameters are calculated for modelling electron-ion recombination fluctuations. The parameters for both the ER and NR cases are functions of a number of (different) underlying nuisance parameters, which would likely be floated in a computation in the same way as the parameters in table \ref{table:floated_postQuanta}.

\begin{table}[H]
\caption{\label{table:yieldParams}Parameters for the FlameNEST yield distribution model functions.}
\centering
    \begin{tabu}{| X[c] | X[c] |}
    \hline
    \textbf{Symbol} & \textbf{Meaning} \\ \hline
    $\overline{n^{\text{el}}} (E)$ & Electron mean yield \\ \hline
    $\overline{n^{\text{q}}} (E)$ & Electron + photon mean yield \\ \hline
    $r_{\text{ex}} (E)$ & Ratio of mean exciton yield to mean ion yield \\ \hline
    $\mathcal{F}_{\text{ER}} (\overline{n^{\text{q}}})$ & ER Fano factor \\ \hline
    $P_{\text{rec}} (\overline{n^{\text{el}}}, \overline{n^{\text{q}}}, r_{\text{ex}})$ & Electron-ion recombination probability \\ \hline
    $\xi (\overline{n^{\text{q}}})$ & Electron-ion recombination skewness parameter \\ \hline
    $\sigma_{\text{rec}} (\overline{n^{\text{el}}}, \overline{n^{\text{q}}}, P_{\text{rec}}, n^{\text{i}}_{\text{prod}})$ & Electron-ion recombination width \\ \hline
    $\delta \sigma (\xi)$ & Electron-ion recombination width correction \\ \hline
    $\delta \mu (\xi, \sigma, \delta \mu)$ & Electron-ion recombination mean correction \\ \hline
    \end{tabu}
\end{table}

\subsection{Post-quanta models}\label{appendix:postquanta}

In this section, we provide the precise post-quanta model descriptions implemented in FlameNEST. Equations \ref{eq:postQuantaS1_first} to \ref{eq:postQuantaS1_last} list the distributions describing the blocks going from produced photons to S1 signal, depicted in the lower row of the post-quanta blocks in Figure \ref{fig:newblockstructure}. It should be noted that the original NEST models perform the final smearing as a two-step process, whereas we use the well-known property of two subsequent normal smearings to model this as a single step, adding the variances in quadrature.

\begin{align}
    P(n^{\text{ph}}_{\text{det}}|n^{\text{ph}}_{\text{prod}}) = \zeta^{\text{ph}}(n^{\text{ph}}_{\text{det}}, \gamma_{\text{min}}) \Binom\left(n^{\text{ph}}_{\text{det}}|n^{\text{ph}}_{\text{prod}},g1 f_{\text{S1}}(r,z)\right) \label{eq:postQuantaS1_first}\\
    P(n^{\text{phel}}_{\text{prod}}|n^{\text{ph}}_{\text{det}}) = \Binom\left(n^{\text{phel}}_{\text{prod}}-n^{\text{ph}}_{\text{det}}|n^{\text{ph}}_{\text{det}}, P_{\text{dpe}}\right) \\
    P(n^{\text{phel}}_{\text{det}}|n^{\text{phel}}_{\text{prod}}) = \Binom\left(n^{\text{phel}}_{\text{det}}|n^{\text{phel}}_{\text{prod}}, P_{\text{spe}}(\mu_{\text{spe}},N_{\text{PMT}})\right) \\
    P(S1|n^{\text{phel}}_{\text{det}}) = \xi^{\text{S1}}(S1, S1_{\text{min}}, S1_{\text{max}}) \Normal\left(S1|n^{\text{phel}}_{\text{det}}, \sqrt{\sigma_{\text{spe}}^2 n^{\text{phel}}_{\text{det}} + \Delta_{\text{S1}}^2 (n^{\text{phel}}_{\text{det}})^2}\right) \label{eq:postQuantaS1_last}
\end{align}

Equations \ref{eq:postQuantaS2_first} to \ref{eq:postQuantaS2_last} list the distributions corresponding to the upper row of post-quanta model blocks in Figure \ref{fig:newblockstructure}, going from produced electrons to S2 signal. The tilde denotes an applied continuity correction, as detailed in the main text.

\begin{align}
    P(n^{\text{el}}_{\text{det}}|n^{\text{el}}_{\text{prod}}) = \Binom\left(n^{\text{el}}_{\text{det}}|n^{\text{el}}_{\text{prod}},\eta^{\text{el}}(z,z_{\text{topDrift}},v_{\text{drift}},\tau,\epsilon_{\text{gas}})\right) \label{eq:postQuantaS2_first} \\
    P(n^{\text{S2-ph}}_{\text{prod}}|n^{\text{el}}_{\text{det}}) = \widetilde{\Normal}\left(n^{\text{S2-ph}}_{\text{prod}}|\mu_{\text{el}}(\epsilon_{\text{gas}},\rho_{\text{gas}},\Delta_{\text{gas}})n^{\text{el}}_{\text{det}}, \sigma_{\text{el}}(\epsilon_{\text{gas}},\rho_{\text{gas}},\Delta_{\text{gas}},\mathcal{F}_{\text{S2}})\sqrt{n^{\text{el}}_{\text{det}}}\right) \\
    P(n^{\text{S2-ph}}_{\text{det}}|n^{\text{S2-ph}}_{\text{prod}}) = \Binom\left(n^{\text{S2-ph}}_{\text{det}}|n^{\text{S2-ph}}_{\text{prod}},g1_{\text{gas}} f_{\text{S2}}(r)\right) \\
    P(n^{\text{S2-phel}}|n^{\text{S2-ph}}_{\text{det}}) = \Binom\left(n^{\text{S2-phel}}-n^{\text{S2-ph}}_{\text{det}}|n^{\text{S2-ph}}_{\text{det}}, P_{\text{dpe}}\right) \\
    P(S2|n^{\text{S2-phel}}) = \xi^{S2}(S2, S2_{\text{min}}, S2_{\text{max}}) \Normal\left(S2|n^{\text{S2-phel}}, \sqrt{\sigma_{\text{spe}}^2 n^{\text{S2-phel}} + \Delta_{\text{S2}}^2 (n^{\text{S2-phel}})^2}\right) \label{eq:postQuantaS2_last}
\end{align}

\subsection{Pre-quanta models}\label{appendix:prequanta}
In this section, we provide the full description of the pre-quanta models implemented in FlameNEST. Equations \ref{eq:preQuantaER_first} - \ref{eq:preQuantaER_last} list the probability distributions used to calculate the pre-quanta model block in the ER case. The tilde denotes an applied continuity correction, whilst the hat denotes the condition $n^{\text{el}}_{\text{prod}} \leq n^{\text{i}}_{\text{prod}}$ discussed in the main text being applied at the level of the distribution. This is detailed more in Appendix \ref{appendixb}.

\begin{align}
    P(n^{\text{q}}_{\text{prod}}|\overline{n^{\text{q}}})_{\text{ER}} = \widetilde{\Normal} \left(n^{\text{q}}_{\text{prod}}|\overline{n^{\text{q}}}, \sqrt{\mathcal{F}_{\text{ER}}\overline{n^{\text{q}}}}\right) \label{eq:preQuantaER_first} \\
    P(n^{\text{i}}_{\text{prod}}|n^{\text{q}}_{\text{prod}})_{\text{ER}} = \Binom \left(n^{\text{i}}_{\text{prod}}|n^{\text{q}}_{\text{prod}}, \alpha \right) \label{eq:preQuantaER_mid} \\
    P(n^{\text{el}}_{\text{prod}}|n^{\text{i}}_{\text{prod}})_{\text{ER}} = \widehat{\SkewNormal} \left(n^{\text{el}}_{\text{prod}}|(1-P_{\text{rec}})n^{\text{i}}_{\text{prod}} - \delta \mu, \frac{\sigma_{\text{rec}}}{\delta \sigma}, \xi \right) \label{eq:preQuantaER_last}
\end{align}

The distributions used to calculate the pre-quanta model block for NR interactions are listed in Equations \ref{eq:preQuantaNR_first} - \ref{eq:preQuantaNR_last}. The tilde and hat take the same meaning as for the ER case.

\begin{align}
    P(n^{\text{i}}_{\text{prod}}|\overline{n^{\text{q}}})_{\text{NR}} = \widetilde{\Normal }\left(n^{\text{i}}_{\text{prod}}|\alpha \overline{n^{\text{q}}}, \sqrt{\alpha \overline{n^{\text{q}}}}\right) \label{eq:preQuantaNR_first} \\
    P(n^{\text{q}}_{\text{prod}}|\overline{n^{\text{q}}}, n^{\text{i}}_{\text{prod}})_{\text{NR}} = \widetilde{\Normal} \left(n^{\text{q}}_{\text{prod}} - n^{\text{i}}_{\text{prod}}|\alpha \overline{n^{\text{q}}} r_{\text{ex}}, \sqrt{\alpha \overline{n^{\text{q}}} r_{\text{ex}}}\right) \\
    P(n^{\text{el}}_{\text{prod}}|n^{\text{i}}_{\text{prod}})_{\text{NR}} = \widehat{\SkewNormal} \left(n^{\text{el}}_{\text{prod}}|(1-P_{\text{rec}})n^{\text{i}}_{\text{prod}} - \delta \mu, \frac{\sigma_{\text{rec}}}{\delta \sigma}, \xi \right) \label{eq:preQuantaNR_last}
\end{align}

\section{Modified skew Gaussian to implement NEST constraint} \label{appendixb}

As discussed in the main text, NEST implements the condition that $n^{\text{el}}_{\text{prod}} \leq n^{\text{i}}_{\text{prod}}$. We account for this in FlameNEST by modifying the skew Gaussian PDF as follows. The PDF for a standard skew Gaussian distribution with mean $\mu$, standard deviation $\sigma$ and skewness parameter $\alpha$ takes the form

\begin{equation}
    f(x;\mu,\sigma,\alpha) = \frac{1}{\sqrt{2\pi\sigma^2}}\exp\left(-\frac{(x-\mu)^2}{2\sigma^2}\right) \left(1+\erf\left[\left(\frac{\alpha}{\sqrt{2}\sigma}\right)(x-\mu)\right]\right).
\end{equation}

In FlameNEST, we modify this to read

\begin{equation}
\label{eq:truncatedSkewGauss}
    f(x;\mu,\sigma,\alpha, l) =
    \left\{
    \begin{array}{ll}
        \frac{1}{\sqrt{2\pi\sigma^2}}\exp\left(-\frac{(x-\mu)^2}{2\sigma^2}\right) \left(1+\erf\left[\left(\frac{\alpha}{\sqrt{2}\sigma}\right)(x-\mu)\right]\right) & x < l \\
        1 - \left\{ \frac{1}{2}\left(1+\erf\left[\frac{x-\mu}{\sqrt{2}\sigma}\right]\right) - 2T\left(\frac{x-\mu}{\sigma},\alpha\right) \right\} & x = l \\
        0 & x > l \\
    \end{array} 
    \right.
\end{equation}

\noindent where $x$ maps to $n^{\text{el}}_{\text{prod}}$ and $l$ to $n^{\text{i}}_{\text{prod}}$. The term in curly brackets in the $x=l$ case is the cumulative distribution function (CDF) of the skew Gaussian distribution, and T is Owen's T function \cite{Owen}. This has the effect of `re-dumping' all probability mass for $x > l$ into the probability mass at $x = l$, once a continuity correction is applied as in Equation \ref{eq:contCorr}, which is an appropriate capturing of NEST's MC behaviour, setting any sampled $n^{\text{el}}_{\text{prod}} > n^{\text{i}}_{\text{prod}}$ to be equal to $n^{\text{i}}_{\text{prod}}$.

Implementing this as a \texttt{TensorFlow} computation required adding a custom distribution to the \texttt{TensorFlow Probability} library \cite{tensorflow_developers_2021_5095721}. Of particular importance was an efficient evaluation of Owen's T function $T(h,a)$, which is the integral

\begin{equation}
    T(h,a) = \frac{1}{2\pi} \int_{0}^{a} \frac{e^{-\frac{1}{2}h^2(1+x^2)}}{1+x^2} dx .
\end{equation}

In our case $a \geq 0$. Owen proved the relation \cite{Owen}

\begin{equation}
    T(h,a) = \frac{1}{2}\Phi(h) + \frac{1}{2}\Phi(ah) - \Phi(h)\Phi(ah) - T\left(ah,\frac{1}{a}\right),
\end{equation}

\noindent where $\Phi$ is the CDF of the standard normal distribution, and so we can always recast $T(h,a)$ to be in $0 \leq a \leq 1$. It is then straightforward to perform a Taylor expansion in $a$

\begin{equation}
    T(h,a) = \frac{1}{2\pi} \left\{ tan^{-1}(a) + \sum_{i=1}^{\infty}C_i\frac{a^{2i-1}}{2i-1} \right\},
\end{equation}

\noindent where the coefficients are obtained recursively as

\begin{equation}
    \begin{split}
        C_1 = e^{-\frac{h^2}{2}} - 1, \\
        C_{n+1} = -C_n + (-1)^n \frac{(\frac{h^2}{2})^n}{n!} e^{-\frac{h^2}{2}}.
    \end{split}
\end{equation}

We determined that in our application of Equation \ref{eq:truncatedSkewGauss} a sufficient degree of accuracy could be obtained for all relevant parameter values with a truncation of the series at $C_2$ for NR sources and $C_5$ for ER sources. This is evident from the results in Sections \ref{subsec:mono_validations} and \ref{subsec:uniform_validations}.

\section{Manual ion bound computation in FlameNEST} \label{appendixc}

As discussed in the main text, for the FLameNEST block structure a manual calculation is done for the ion bounds, constructing different bounds for each energy summed over in the quanta tensor. In the ER case, the following quantities are first calculated, representing bounds on $n^q_{\text{prod}}$, coming from distribution in Equation \ref{eq:preQuantaER_first},

\begin{align}
    n^q_{\text{upper}} = \overline{n^q} + \sigma \sqrt{\mathcal{F} \overline{n^q}} \\
    n^q_{\text{lower}} = \overline{n^q} - \sigma \sqrt{\mathcal{F} \overline{n^q}}.
\end{align}

All symbols have the same meaning as in Appendix \ref{appendix:params}, and $\sigma$ is a user-defined parameter controlling the width of the bounds. It should be noted that energy enters implicitely in $\overline{n^q}$. Upper and lower bounds on the mean and standard deviation of the number of ions described by the binomial of Equation \ref{eq:preQuantaER_mid} are then calculated as

\begin{align}
    \mu_{\text{upper/lower}} = n^q_{\text{upper/lower}} \alpha \\
    \sigma_{\text{upper/lower}} = \sqrt{n^q_{\text{upper/lower}} \alpha (1-\alpha)}.
\end{align}

In the NR case, the upper and lower bounds on the mean and standard deviation of the number of ions described by the normal distribution of Equation \ref{eq:preQuantaNR_first} can simply be calculated as

\begin{align}
    \mu_{\text{upper}} = \mu_{\text{lower}} = \overline{n^q} \alpha \\
    \sigma_{\text{upper}} = \sigma_{\text{lower}} = \sqrt{\overline{n^q} \alpha}.
\end{align}

Then, upper and lower bounds on the number of ions can be calculated straightforwardly as

\begin{align}
    n^i_{\text{min}} = \mu_{\text{lower}} - \sigma \sigma_{\text{lower}} \\
    n^i_{\text{max}} = \mu_{\text{upper}} + \sigma \sigma_{\text{upper}}.
\end{align}

\end{document}